\documentclass[%
 reprint,
 superscriptaddress,
 amsmath,amssymb,
 aps,
 prl,
longbibliography
]{revtex4-1}

\usepackage{graphicx}
\usepackage{dcolumn}
\usepackage{xcolor}
\usepackage{bm}

\usepackage[colorlinks=true,linkcolor=blue, citecolor=blue]{hyperref}%

\begin{document}

\title{Coupled Spin-Orbital $p$-Wave Magnetism via Structural and Magnetic Chirality}

\author{Tom G. Saunderson}
\email{thomas.saunderson@physik.uni-halle.de}
\affiliation{Institut f\"ur Physik and Halle-Berlin-Regensburg Cluster of Excellence CCE, Martin-Luther-Universit\"at Halle-Wittenberg, D-06099 Halle (Saale), Germany} 
\author{B\"orge G\"obel}
\affiliation{Institut f\"ur Physik and Halle-Berlin-Regensburg Cluster of Excellence CCE, Martin-Luther-Universit\"at Halle-Wittenberg, D-06099 Halle (Saale), Germany}
\author{Ersoy \c{S}a\c{s}\i o\u{g}lu}
\affiliation{Institut f\"ur Physik and Halle-Berlin-Regensburg Cluster of Excellence CCE, Martin-Luther-Universit\"at Halle-Wittenberg, D-06099 Halle (Saale), Germany}
\author{Samir Lounis}%
\affiliation{Institut f\"ur Physik and Halle-Berlin-Regensburg Cluster of Excellence CCE, Martin-Luther-Universit\"at Halle-Wittenberg, D-06099 Halle (Saale), Germany}

\date{\today}

\begin{abstract}

Helical spin textures represent the minimal realization of $p$-wave magnetism which is characterized by momentum-odd spin polarization. Independently, structurally chiral crystals exhibit momentum-odd orbital polarization arising from broken inversion symmetry. Here, we demonstrate that spin-orbit coupling couples these two independent microscopic chirality degrees of freedom, allowing the orbital polarization of a chiral crystal to generate an additional contribution to the $p$-wave spin splitting. The resulting spin-orbital state is naturally classified by the relative chirality $\eta=\chi_{\mathrm c}\chi_{\mathrm m}$, giving rise to two symmetry-distinct $p$-wave phases corresponding to homochiral and heterochiral configurations which can be directly probed by the longitudinal conductivity. These phases exhibit distinct transport signatures, establishing a unified framework linking orbitronics and unconventional magnetism through coupled spin-orbital $p$-wave order.

\end{abstract}

\maketitle


\begin{figure*}[t!]
\includegraphics*[width=0.9\linewidth,clip]{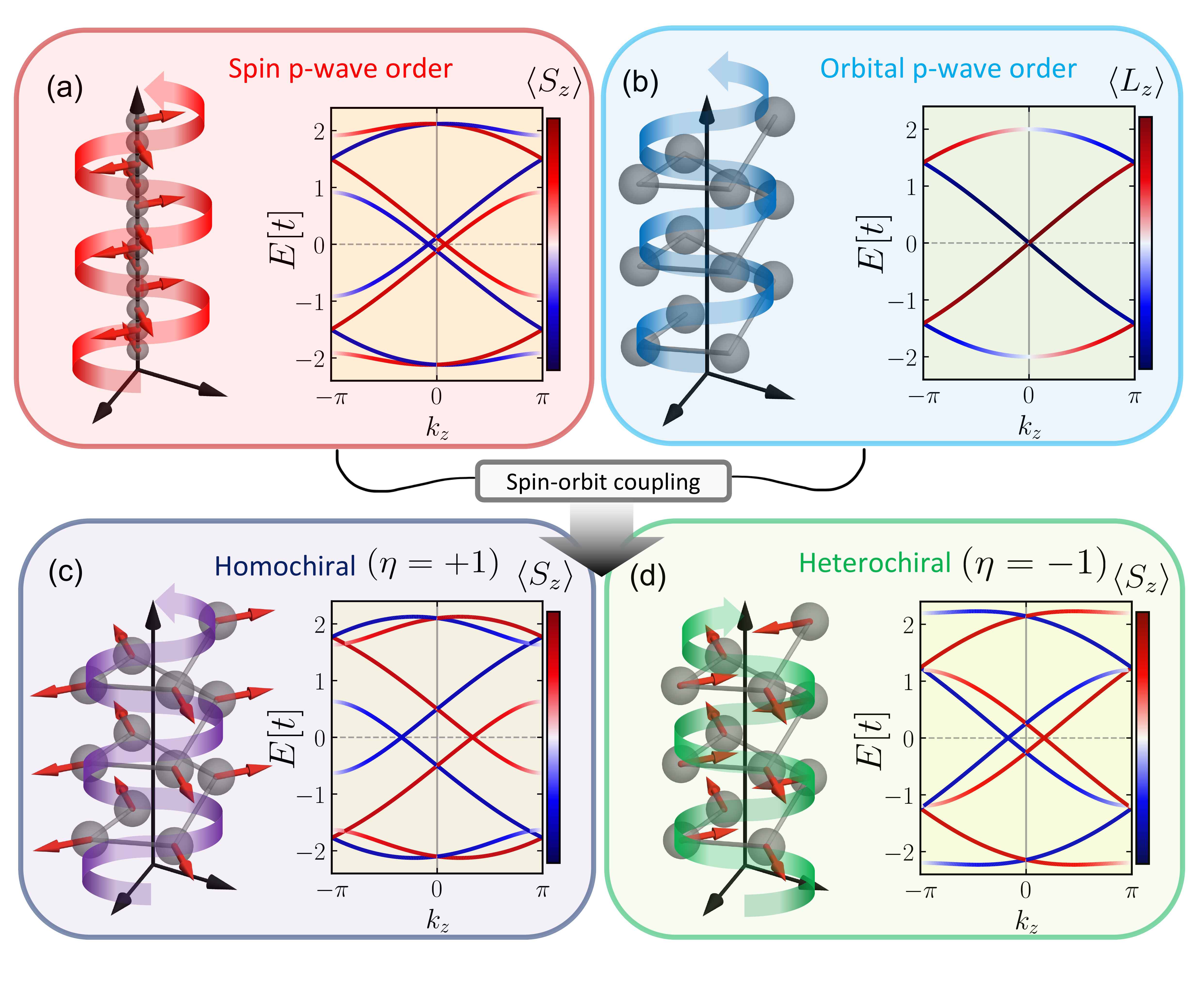}
\caption{
\textbf{Magnetic and structural chirality generate coupled spin-orbital $p$-wave order.}
(a) One-dimensional helical spin spiral exhibiting momentum-odd spin polarization characteristic of magnetic $p$-wave order.
(b) Structurally chiral one-dimensional wire generating momentum-odd orbital polarization, establishing structural chirality as an orbital analogue of $p$-wave order.
(c,d) Spin expectation values for left- and right-handed helical spin spirals on a right-handed chiral chain. Spin-orbit coupling transfers the orbital polarization into the spin sector, where it combines with the magnetic contribution. Depending on the relative handedness of the magnetic and structural chirality, the two $p$-wave channels couple, producing distinct homochiral and heterochiral phases that fundamentally alter the symmetry of the underlying electronic structure.
}
\label{fig:Fig1}
\end{figure*}

The discovery of unconventional spin-polarized states protected by crystal and magnetic symmetries has fundamentally broadened the classification of magnetic order beyond conventional ferromagnetism. While ferromagnets exhibit exchange-driven $s$-wave spin splitting that preserves its sign throughout momentum space, compensated magnetic systems have recently been shown to support higher-order momentum-dependent spin textures analogous to the orbital classification of atomic wavefunctions and the pairing symmetries of unconventional superconductors~\cite{Gonzalez-Hernandez2021,Smejkal2022,Smejkal2022a}. The realization of altermagnetism demonstrated that compensated magnets can host large momentum-dependent spin splitting without spin-orbit coupling, establishing symmetry as a powerful route towards unconventional magnetic phases~\cite{Smejkal2022,Mazin2024PRX,Amin2024a,Fedchenko2024}. Extracting pure spin currents from even-parity altermagnets generally relies on momentum-selective transport or carefully engineered interfaces that filter opposite spin channels~\cite{Gonzalez-Hernandez2021,Fu2025}. More recently, rolling d-wave magnets into nanotubes has been proposed as a novel route to pure spin-current generation~\cite{asoasoGlu2026}.

More recently, this classification has been extended to odd-parity magnetic phases. Helical magnetic textures were shown to generate momentum-odd spin polarization, establishing the concept of \emph{p}-wave magnetism in which the spin polarization reverses sign under momentum inversion while remaining fully compensated in real space~\cite{Hellenes2024,Brekke2024}. Subsequently, the concept was generalized to collinear compensated magnets through mirror-asymmetric magnetic motifs and magnetoelectric coupling, demonstrating that odd-parity spin splitting is not restricted to non-collinear magnetism~\cite{Cardenas2026}. Very recently, electrical switching and experimental observation of p-wave magnetism have established odd-parity magnetic phases as experimentally accessible states of matter, motivating the search for additional mechanisms capable of generating momentum-odd electronic polarization~\cite{Song2025,Chakraborty2025,McNally2026,Zhu2026}. Furthermore, rolling 2-dimensional materials into nanotubes has already enabled forms of 1-dimensional p-wave magnetism \cite{Jin2026}, while helical chains represent its minimal realisation \cite{Gobel2026}. Collectively, these developments establish p-wave magnetism as an emerging member of the broader family of unconventional symmetry-protected magnetic phases.

Independently, structurally chiral crystals have emerged as a rich platform for momentum-dependent orbital phenomena \cite{Hagiwara2025,Go2018a,Go2021c}. Broken inversion symmetry in chiral crystals gives rise to orbital textures and orbital Edelstein responses through orbital magnetization, even in systems without explicit atomic orbital degrees of freedom~\cite{Gobel2025,Ceresoli2006,Xiao2010,Shi2007}. The same nanotube geometry also gives rise to momentum-dependent orbital textures, providing the orbital analogue of the spin phenomena discussed above~\cite{Yoda2015,Gobel2025a}. These orbital textures possess odd parity in momentum space and therefore exhibit the same fundamental symmetry as the momentum-dependent spin polarization characteristic of p-wave magnetism. Despite this close correspondence, orbital chirality and p-wave magnetism have so far been developed as separate research directions.

The coexistence of momentum-odd spin and orbital polarization raises a fundamental question: can structural chirality itself realize an orbital analogue of $p$-wave magnetism? More generally, can structural and magnetic chirality be understood as two independent microscopic manifestations of a common odd-parity order parameter? Since spin-orbit coupling directly couples orbital and spin degrees of freedom, chiral magnetic systems provide a natural setting in which these two momentum-odd channels may coexist and interact.

In this work we demonstrate that structural and magnetic chirality constitute two independent microscopic symmetry degrees of freedom. Spin-orbit coupling couples these two chiralities, giving rise to two symmetry-distinct spin-orbital $p$-wave phases corresponding to homochiral and heterochiral configurations. Using a tight-binding model of a structurally chiral helix with compensated helical magnetic order, we show that the orbital polarization generated by structural chirality produces an additional contribution to the $p$-wave spin splitting. Our results establish relative chirality as the fundamental symmetry descriptor of coupled spin-orbital $p$-wave order.

To isolate the minimal ingredients required for coupled spin-orbital p-wave order, we employ the analytically solvable tight-binding model \cite{Gobel2025} for structurally chiral one-dimensional helices. The model consists of a helical chain with a single $s$ orbital per atomic site, in which structural chirality is encoded entirely through the geometry of the hopping network. The orbital magnetization is evaluated using the modern theory of orbital magnetization, naturally giving rise to a momentum-odd orbital polarization despite the absence of atomic orbital degrees of freedom. Next we introduce a helical spin spiral by a local exchange field \cite{Gobel2026}, while spin-orbit coupling is incorporated via an effective spin-dependent hopping following Ref.~\cite{Gobel2025}. Unless otherwise stated, all calculations presented below are performed using this Hamiltonian, details of which are provided in the Supplementary \cite{Supplementary}.


We begin by considering a one-dimensional helical spin chain, illustrated schematically in Fig.~\ref{fig:Fig1}(a). The continuously rotating magnetic texture possesses a well-defined handedness and generates a momentum-dependent spin polarization in the electronic structure. Previous studies have shown that such helical magnetic states exhibit the characteristic odd-parity spin symmetry associated with p-wave magnetism, whereby the spin polarization reverses under momentum inversion,
\begin{equation}
\langle S_z(k)\rangle = -\langle S_z(-k)\rangle
\label{eq:spinodd}
\end{equation}
the essential ingredient is the chirality of the magnetic texture itself. This form of chirality is denoted by the quantity $\chi_m= \pm1$ whose sign follows the handedness of the spiral. Reversing the handedness of the spin spiral reverses the sign of the momentum-dependent spin polarization, and $\chi_m$, establishing a direct correspondence between magnetic chirality and p-wave spin order.

The existence of momentum-odd orbital polarization immediately suggests that structural chirality, $\chi_c= \pm1$, may constitute an orbital analogue of $p$-wave magnetism. We now consider the corresponding orbital analogue shown in Fig.~\ref{fig:Fig1}(b). In this case the chirality originates not from the magnetic texture but from the crystal structure itself. Such systems naturally generate momentum-dependent orbital polarization, producing orbital textures that are odd under momentum inversion,
\begin{equation}
\langle L_z(k)\rangle = -\langle L_z(-k)\rangle.
\label{eq:orbital_odd}
\end{equation}
The striking observation is that the orbital texture possesses the same symmetry as the spin texture of a helical magnetic chain. In both cases, the handedness of the underlying chiral object determines the sign of a momentum-odd polarization. Structural chirality therefore provides an orbital analogue of $p$-wave magnetic order, generating a momentum-dependent orbital polarization without requiring magnetic order.

\begin{figure}[h]
\includegraphics*[width=0.9\linewidth,clip]{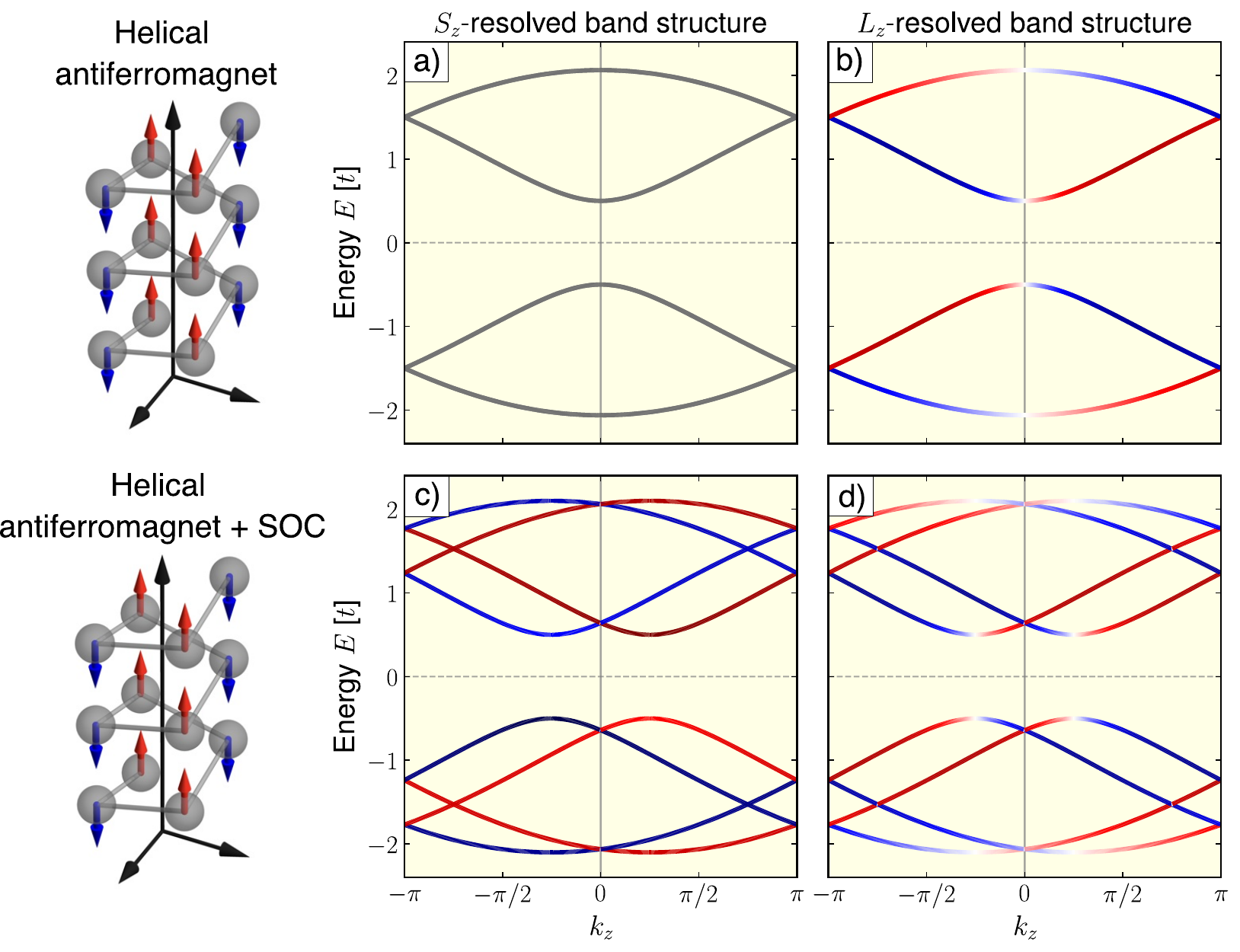}
\caption{
\textbf{Spin-orbit coupling transfers orbital $p$-wave order into the spin sector.}
Compensated collinear antiferromagnetic order on a structurally chiral one-dimensional wire. (a) Spin-resolved and (b) orbital-resolved band structures in the absence of spin-orbit coupling. Although the collinear antiferromagnetic texture exhibits no momentum-dependent spin splitting, the structurally chiral lattice generates momentum-odd orbital polarization. (c) Spin-resolved and (d) orbital-resolved band structures in the presence of spin-orbit coupling. The orbital $p$-wave polarization is transferred into the spin sector, inducing momentum-dependent spin splitting while preserving the orbital texture. This limiting case demonstrates that structural chirality alone provides an independent microscopic route to spin $p$-wave magnetism.
}
\label{fig:Orbital_pwave_magnet}
\end{figure}

This orbital $p$-wave order survives in the case of a collinear antiferromagnet, shown in Fig.~\ref{fig:Orbital_pwave_magnet}, thereby constituting the limiting case of an orbital p-wave magnet. In the absence of spin-orbit coupling, the antiferromagnetic texture exhibits no momentum-dependent spin splitting despite retaining the momentum-odd orbital polarization generated by the structurally chiral lattice. Upon introducing spin-orbit coupling, the orbital polarization is transferred into the spin sector, inducing a momentum-dependent spin splitting. This limiting case demonstrates that structural chirality provides an independent microscopic route to spin $p$-wave magnetism.


The coexistence of these two odd-parity polarization channels immediately raises the central question of this work: do spin and orbital p-wave order remain independent, or do they combine to generate entirely new spin-orbital phases? While the orbital polarization remains confined to the orbital sector in the absence of relativistic interactions, spin-orbit coupling provides a natural mechanism through which the orbital chirality may be projected onto the spin degree of freedom. The resulting system contains two independent p-wave channels, one originating from magnetic chirality and the other from structural chirality, whose interplay forms the central focus of this work.

\begin{figure}[h]
\includegraphics*[width=0.9\linewidth,clip]{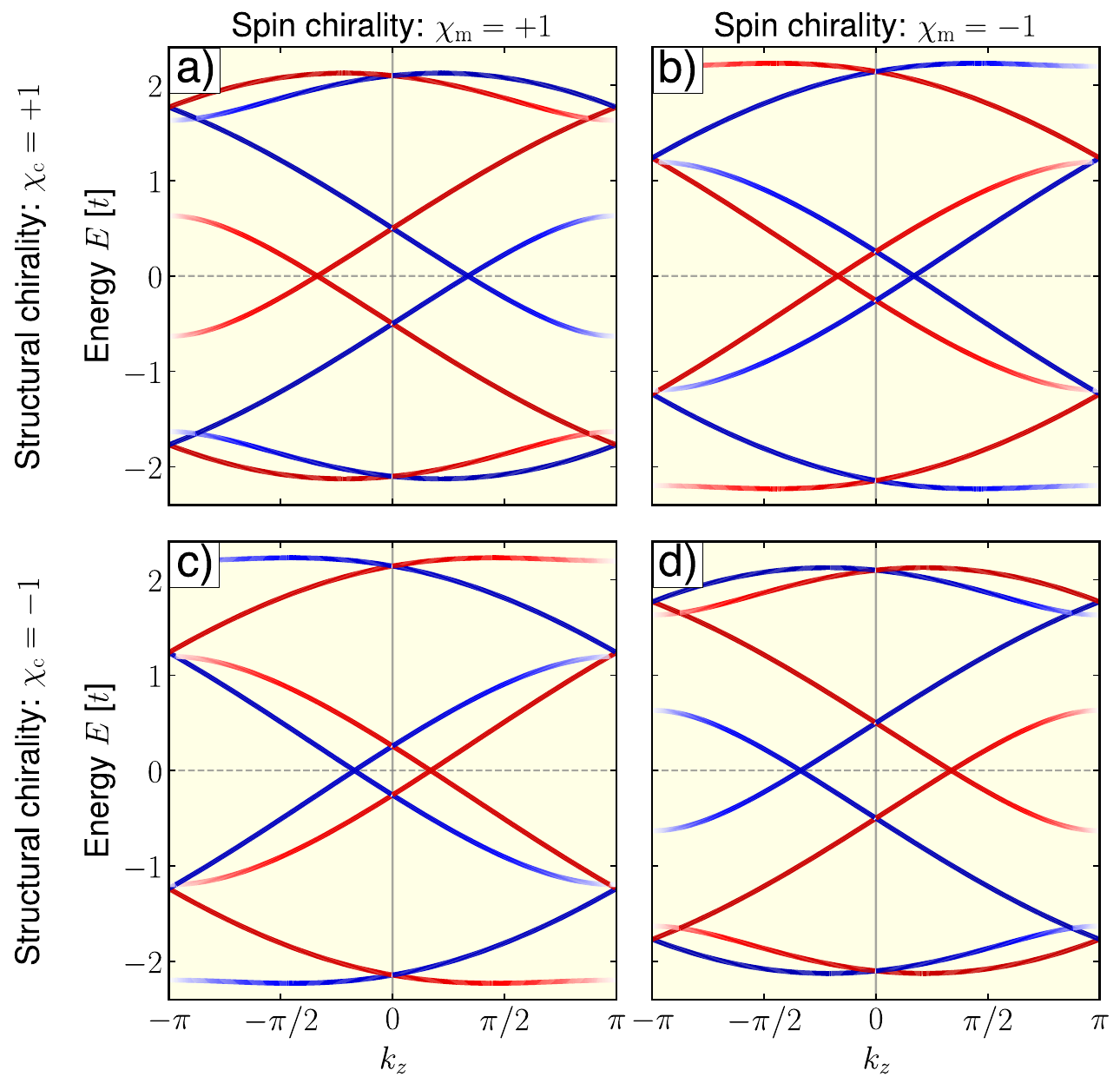}
\caption{
\textbf{Relative chirality defines distinct spin-orbital $p$-wave phases.}
Spin-polarized band structures for (a) a left-handed spin helix on a left-handed chiral chain, (b) a right-handed spin helix on a left-handed chiral chain, (c) a left-handed spin helix on a right-handed chiral chain, and (d) a right-handed spin helix on a right-handed chiral chain. The relative handedness of the magnetic and structural chirality determine whether the combined state enters the homochiral or heterochiral magnetic phases. Configurations with matching handedness [(a,d)] form the homochiral spin-orbital $p$-wave phase, whereas opposite-handed configurations [(b,c)] form the heterochiral phase. These two symmetry sectors exhibit distinct momentum-dependent spin textures and electronic structures.
}
\label{fig:LHRHspin_vs_LHRHorbital}
\end{figure}

\begin{figure}[h]
\includegraphics*[width=0.9\linewidth,clip]{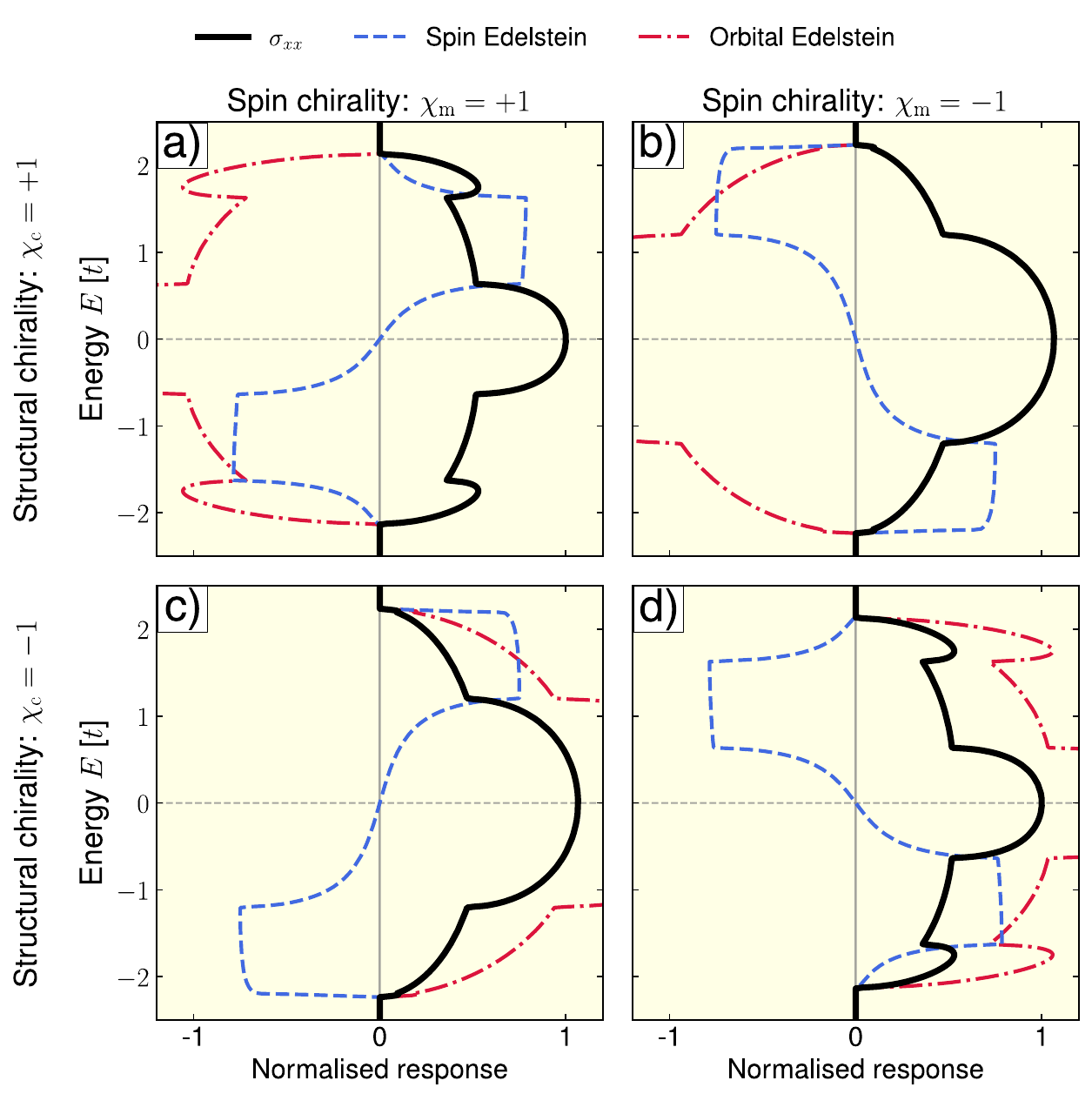}
\caption{\textbf{Transport signatures of coupled spin-orbital \textit{p}-wave order.}
Normalized longitudinal conductivity $\sigma_{xx}$ (solid black), spin Edelstein response (dashed blue), and orbital Edelstein response (dash-dotted red) as functions of the Fermi energy for the four combinations of structural and magnetic chirality: (a) right-handed crystal/right-handed helix, (b) right-handed crystal/left-handed helix, (c) left-handed crystal/right-handed helix, and (d) left-handed crystal/left-handed helix. All responses are normalized to $\sigma_{xx}^{\mathrm{LH,LH}}(E_F)$, indicated by the vertical grey line. Reversing the structural chirality reverses the orbital Edelstein response, while reversing the magnetic chirality reverses the spin Edelstein response, reflecting the independent orbital and spin \textit{p}-wave channels. In contrast, the longitudinal conductivity is insensitive to these symmetry-imposed sign reversals and instead reveals changes in the underlying electronic structure arising from the coupling between structural and magnetic chirality. The conductivity therefore provides a direct experimental signature of the coupled spin-orbital \textit{p}-wave phase by distinguishing the homochiral [(a,d)] and heterochiral [(b,c)] sectors through characteristic changes in transport magnitude.
}
\label{fig:Transport}
\end{figure}

We now consider the coexistence of structural and magnetic chirality in
the presence of spin-orbit coupling. The combined system is naturally described by the structural and magnetic
chirality indices $(\chi_{\mathrm c},\chi_{\mathrm m})$, giving four
possible microscopic configurations.

Figure~\ref{fig:LHRHspin_vs_LHRHorbital} presents the corresponding
spin-polarized band structures. These four microscopic configurations
collapse into two symmetry-distinct sectors classified by the relative
chirality
\begin{equation}
\eta=\chi_{\mathrm c}\chi_{\mathrm m}.
\end{equation}
Configurations with $\eta=+1$ form the \emph{homochiral} phase, while
those with $\eta=-1$ form the \emph{heterochiral} phase. Within each
sector the two microscopic configurations are related by the inversion of the spin polarization, whereas the
homochiral and heterochiral sectors exhibit distinct electronic
structures and momentum-dependent spin textures.


The existence of these distinct spin-orbital phases is directly reflected in their transport signatures, shown in Fig.~\ref{fig:Transport}. When an electric field is applied along the periodic direction a longitudinal current and non-equilibrium spin and orbital moments are generated. We quantify these responses to the field with the longitudional conductivity, $\sigma_{zz}$, spin $\chi^{S_z}$ and orbital $\chi^{L_z}$ susceptibilities. Reversing the structural chirality reverses only the orbital Edelstein response [e.g. Figs.~\ref{fig:Transport}(a to c)], while reversing the magnetic chirality reverses only the spin Edelstein response [e.g. Figs.~\ref{fig:Transport}(a to b)], demonstrating that the orbital and spin p-wave channels remain independently observable through their respective transport responses. Beyond these symmetry-imposed sign reversals, systematic changes in the magnitudes of both responses reveal the electronic reconstruction associated with the homochiral and heterochiral phases identified in Fig.~\ref{fig:LHRHspin_vs_LHRHorbital}.

This electronic reconstruction is most clearly revealed by the longitudinal charge conductivity, whose sign is unaffected by reversing either chirality. Unlike the spin and orbital Edelstein responses, which directly identify the reversal of the individual p-wave channels, the longitudinal conductivity isolates the electronic reconstruction arising from their coupling. It therefore provides the clearest experimental signature of the coupled spin-orbital p-wave phase, distinguishing the homochiral and heterochiral phases through characteristic changes in transport magnitude.

Finally, we discuss potential material realizations of the coupled spin-orbital $p$-wave phase. Chiral B20 helimagnets provide a particularly promising platform owing to the coexistence of structural chirality and long-wavelength helical magnetic order. Experimentally, chemical substitution can tune the magnetic helicity by modifying the Dzyaloshinskii–Moriya interaction and the helical pitch while preserving the crystal chirality, as demonstrated in both the $\beta$-Mn Co--Zn--Mn family and the B20 alloy Mn$_{1-x}$Fe$_x$Ge~\cite{Karube2018,Tokunaga2015,Grigoriev2013,Koretsune2015,Shibata2013}. In particular, Mn$_{1-x}$Fe$_x$Ge undergoes a reversal of magnetic helicity as a function of composition due to a sign change of the effective Dzyaloshinskii–Moriya interaction~\cite{Grigoriev2013,Koretsune2015}. Within the present framework, this directly switches the relative chirality $\eta$ between the homochiral and heterochiral spin-orbital $p$-wave phases, providing an experimentally accessible route to probe the coupling between structural and magnetic chirality. 


In summary, we have shown that structural and magnetic chirality constitute two independent microscopic symmetry degrees of freedom that generate orbital and spin $p$-wave order, respectively. Spin-orbit coupling couples these two chiralities, naturally classifying the resulting spin-orbital state by the relative chirality $\eta=\chi_{\mathrm c}\chi_{\mathrm m}$ and giving rise to homochiral and heterochiral $p$-wave phases with distinct electronic structures and transport signatures.

These findings establish a unified framework linking orbitronics and unconventional magnetism through coupled spin-orbital p-wave order, identifying relative chirality as a new symmetry degree of freedom in chiral magnetic systems.

\begin{acknowledgments}
The authors acknowledge financial support from the German Excellence Strategy – EXC 3112/1 – 533767171 (Center for Chiral Electronics). T.G.S. acknowledges computing time provided through the Gauss Centre for Supercomputing e.V. on the supercomputer JUWELS at Forschungszentrum Jülich under project superorb.
\end{acknowledgments}


%


\end{document}